\documentclass[aps,prper,reprint,superscriptaddress]{revtex4-2}
\usepackage{graphicx}
\usepackage{amsmath,amssymb}
\usepackage{xcolor}
\usepackage[colorlinks,citecolor=teal,linkcolor=red,urlcolor=magenta,unicode=true]{hyperref}

\begin{document}
	
    \title{Bridging the Gap Between Virtual and Physical Laboratories: A Web-Based Interactive Platform for Undergraduate Physics Practicals}
	
	\author{Ashadul Halder}
	\email{ashadul.halder@gmail.com}
	\affiliation{Department of Physics, St. Xavier's College, \\30, Mother Teresa Sarani, Kolkata-700016, India.}
	
	\author{Shibaji Banerjee}
	\email{shiva@sxccal.edu}
	\affiliation{Department of Physics, St. Xavier's College, \\30, Mother Teresa Sarani, Kolkata-700016, India.}
	
    \begin{abstract}
        The COVID-19 pandemic highlighted the challenges of maintaining hands-on laboratory instruction in undergraduate physics education. In response, we developed and deployed an interactive online physics laboratory platform designed to closely replicate the experimental setups available and curriculum of St. Xavier's College (Autonomous), Kolkata. The platform is designed to closely replicate real experimental arrangements and is aligned with the curriculum, allowing students to prepare effectively before performing physical experiments. Student feedback revealed that 100\% of the respondents rated the platform as beneficial (rating 4 or 5 on a 5-point scale) to improve conceptual understanding and increase confidence in conducting physical experiments. Furthermore, all students agreed that having access to the online prelab simulations is advantageous and recommended its regular use.  These findings highlight the effectiveness of web-based simulations as a complementary and sustainable resource for physics education.
    \end{abstract}
	
    \maketitle
	\onecolumngrid
    \section{Introduction}

        Laboratory courses are an essential part of undergraduate physics education that helps students improve theoretical concepts and develop essential experimental skills through hands-on activities~\cite{hofstein2004}. In addition, it cultivates procedural knowledge in addition to critical thinking, error analysis, and an appreciation of the complexities of real-world measurement. However, access to physical laboratories is not always possible due to various factors. Many institutions face persistent challenges such as limited equipment, high student-to-apparatus ratios and ongoing maintenance demands.
        
        The COVID-19 pandemic is an example of the vulnerability of traditional laboratory instruction. The abrupt transition to remote learning forced educators around the world to look for alternative approaches to practical science education~\cite{mcateer2020}. Reviewing the situation post facto has enabled us to evaluate the situation from some distance and detachment necessary for analysis and presentation. In this context, virtual laboratories and on-line simulations have gained significant importance. Numerous studies indicate that well-designed virtual laboratories can effectively support learning by offering safe, repeatable, and interactive environments to study~\cite{wieman2008, dejong2013}. These tools are essentially grounded in constructivist learning theory, which emphasizes the active construction of knowledge through direct interaction with phenomena.
        
        Moreover, the use of pre-laboratory simulations has been shown to reduce the students' cognitive load, as described in Cognitive Load Theory~\cite{abdulwahed2011}. By allowing students to become familiar with procedures and equipment beforehand, such tools can lower the barrier to meaningful engagement during actual laboratory sessions. This dual exposure to virtual and physical settings has been associated with improved conceptual understanding and experimental readiness.
        
        Despite these advantages, several limitations persist. There exist many excellent platforms, such as PhET Interactive Simulations~\cite{phet} or the Open Source Physics~\cite{osp} which are geared more toward general conceptual understanding but, at the same time, lack somewhat the procedural fidelity required for university-level experiments~\cite{darrah2014}. Some of the simulations intended function is to serve primarily as demonstrations rather than an actual laboratory environments, reducing opportunities for experiential learning~\cite{brinson2015}. Also, generic simulations often do not align with the specific apparatus, procedures, or evaluation frameworks of individual institutions, which can leave students under-prepared for real laboratory work. In this aspect, the Amrita Virtual Labs \cite{amrita-virtual-labs} and the Virtual Labs initiative by the Ministry of Education \cite{virtual-labs-india} are prominent examples which serve a much wider community of students employing resources and manpower which were beyond our scope at the time of the national emergency and arose as a sort of a spontaneous trigger response to serve this much tighter community of students and teachers. As a consequence, unlike those large platforms, our simulations are built using only HTML, CSS and JavaScript. Users are able to run our simulations completely in the user's browser without needing any server, even internet connection is not required after the page is loaded completely. Users can also download the simulations as simple HTML files along with some extra CSS and JS files, making them easy to use anywhere, even without internet. This makes our platform lightweight, portable and suitable for low-resource settings.
        
        Recent literature emphasizes that authenticity, realism, and curriculum alignment are essential to the effectiveness of virtual laboratories~\cite{smith2024, garcia2024, neovation2023}. Simulations that replicate actual laboratory conditions, allowing users to manipulate variables, collect data manually, and reflect on experimental procedures, are more likely to produce meaningful learning gains. Although they may not correspond to direct hands on learning, it takes a lot less time to check out various options \textit{non-destructively} on the simulator compared to actual lab-time.
        
        To address these challenges, we developed \href{https://openphys.in}{\texttt{https://openphys.in}}, an interactive online physics laboratory platform designed specifically for undergraduate students at the University of Calcutta. The platform closely replicates the real-world experimental setups used in the institution's laboratories and offers an intuitive, browser-based interface for students to engage with core physics experiments. By enabling students to practice procedures, adjust parameters, and analyze simulated data, \href{https://openphys.in}{\texttt{https://openphys.in}} serves as a pedagogical bridge between virtual learning and hands-on experimentation.
        
        This paper evaluates the platform's impact on students' laboratory preparedness, focusing on the development of experimental skills, reduction of cognitive overload, and enhancement of conceptual understanding. Our findings contribute to the broader discourse on digital pedagogy by demonstrating the value of institution-specific, web-based simulations as a sustainable complement to traditional laboratory instruction.

        This paper is organized as follows. \autoref{sec:platform_design} describes the design philosophy, system architecture, and key features of the \texttt{openphys.in} platform. \autoref{sec:res_method} outlines the research methodology used to evaluate its educational impact. In \autoref{sec:result_discus}, the results obtained from student feedback and performance assessments are presented and then analyzed to assess the educational impact of the platform. In \autoref{sec:comp_study}, the strengths of the platform are discussed, along with areas that could be improved. Eventually, a summary of the key findings and the future directions is illustrated in \autoref{sec:conclu}.

    \section{\label{sec:platform_design} Platform Design and Features}

        The \href{https://openphys.in}{\texttt{https://openphys.in}} platform was conceived to address the need for accessible, realistic, and curriculum-aligned virtual laboratory experiences in undergraduate physics education. Its design is grounded in principles of openness, modularity, and ease of deployment, ensuring broad accessibility and adaptability across diverse educational settings \citep{ontodesign,sciencedirectplatform}. The platform supports both independent student learning and integration into formal coursework, making it a robust solution for bridging the gap between virtual and physical laboratory instruction.
        
        \subsection{Design Philosophy and Architecture}
        
        A key consideration in the platform's development was to maximize accessibility and minimize technical barriers for both students and instructors. To this end, \texttt{openphys.in} is implemented entirely using HTML, CSS, and JavaScript, with all simulation logic and user interface components running client-side in the browser. This lightweight, serverless approach eliminates the need for backend runtime environments or complex installations, enabling seamless hosting on static file services such as GitHub\textsuperscript{\textregistered{}} Pages. As a result, the platform can be accessed from any modern web browser without additional software, ensuring high availability and ease of maintenance.
        
        The platform is organized into two main modules: the General Physics Laboratory (GELab) and the Optics Laboratory (OpticsLab). Each module is designed to replicate the structure and content of traditional undergraduate laboratory courses, providing students with a familiar and intuitive virtual environment.
        
        \subsection{General Physics Laboratory (GELab)}
        
        The GELab module focuses on fundamental physics experiments, allowing students to explore core concepts through the following simulations:
        \begin{itemize}
            \item \textbf{Young's Modulus}: Investigate the elastic properties of materials by measuring their ability to withstand length changes under tension or compression.
            \item \textbf{Rigidity Modulus and Moment of Inertia}: Understand the relationship between torque and angular deformation, and study how mass distribution affects rotational motion.
            \item \textbf{Surface Tension}: Examine the cohesive forces at the surface of a liquid that allow it to resist external force.
            \item \textbf{Coefficient of Viscosity}: Explore the internal friction in fluids that affects their flow characteristics.
            \item \textbf{Temperature Coefficient of Resistance}: Analyze how the electrical resistance of materials changes with temperature variations.
            \item \textbf{Stefan's Law}: Delve into the relationship between the temperature of a blackbody and the total energy it emits.
        \end{itemize}
        
        \subsection{Optics Laboratory (OpticsLab)}
        
        The OpticsLab module provides simulations related to light and its interactions, offering experiments such as:
        \begin{itemize}
            \item \textbf{Newton's Rings}: Observe the interference pattern created by the reflection of light between two surfaces—a spherical and a flat surface.
            \item \textbf{Biprism Experiment}: Study the interference of light waves from a single source split into two coherent sources using a biprism.
            \item \textbf{Diffraction Grating}: Analyze the diffraction patterns produced when light passes through a grating with multiple slits, aiding in the measurement of wavelengths.
            \item \textbf{$\mu$-$\lambda$ (Refractive Index and Wavelength) Determination}: Determine the refractive index of materials and the wavelength of light using various optical setups.
            \item \textbf{$i$-$\Delta$ (Angle of Incidence vs. Deviation) Study}: Explore how the angle of incidence affects the deviation of light passing through different media.
        \end{itemize}
        
        \subsection{User Interface and Learning Features}

\begin{figure}[h]
            \centering
            \begin{tabular}{c}
                \includegraphics[width=0.6\linewidth]{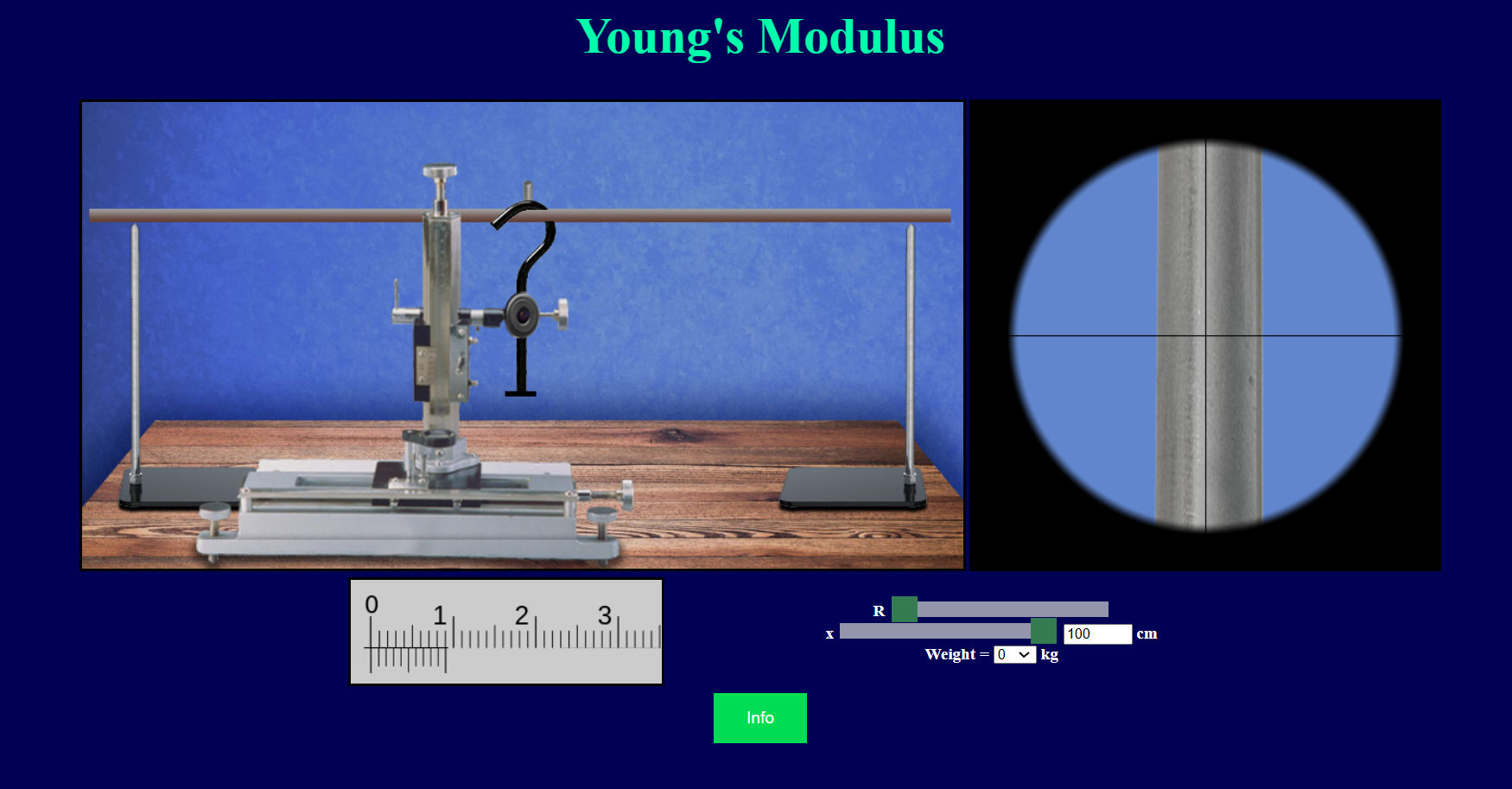}\\
                (a)\\
                \includegraphics[width=0.6\linewidth]{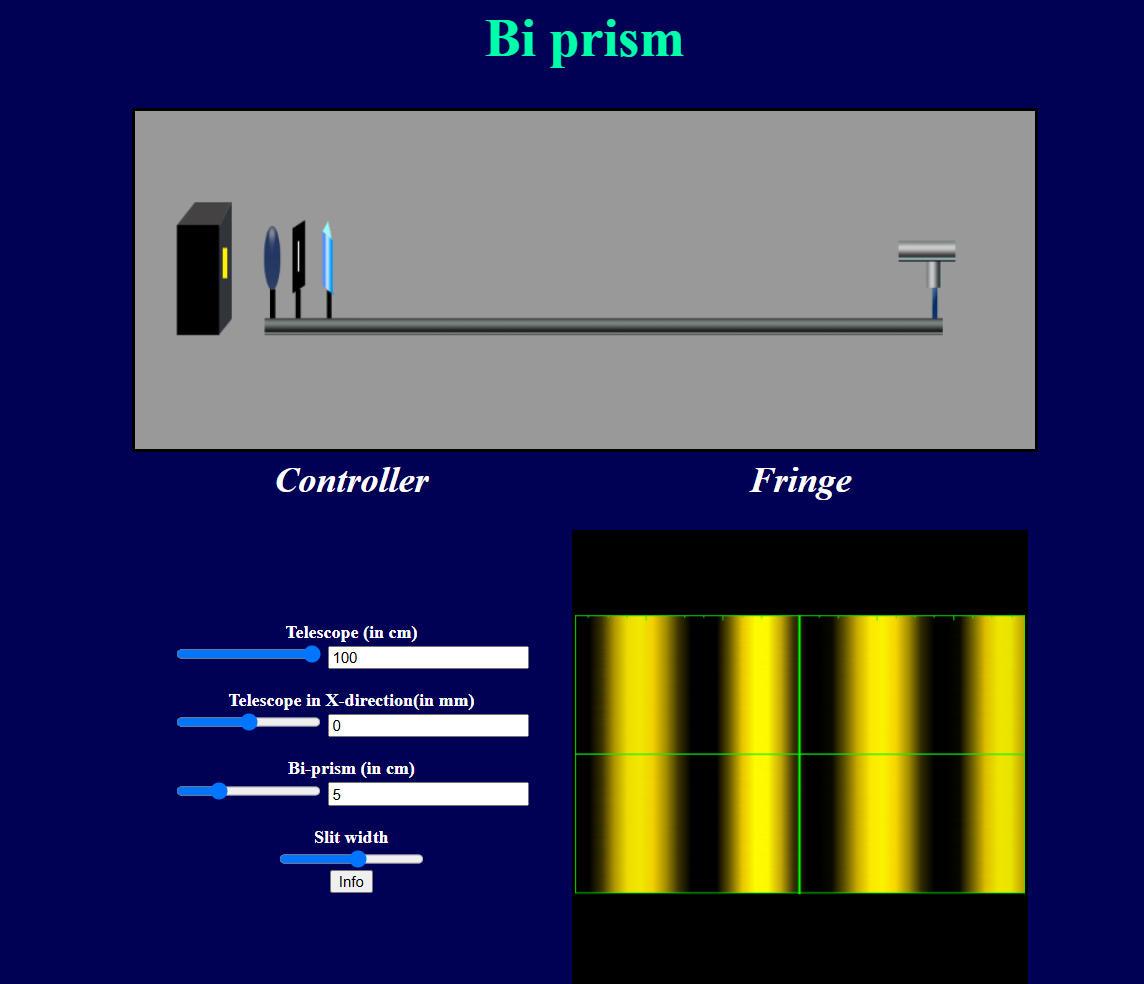}\\
                (b)\\
            \end{tabular}
            \caption{\label{fig:lab_interface} Example interfaces from the online physics laboratory platform available at \href{https://openphys.in/gelab}{\texttt{https://openphys.in/gelab}} and \href{https://openphys.in/opticslab}{\texttt{https://openphys.in/opticslab}}. (a) Young's Modulus experiment from the General Physics Laboratory (GELab), which features a visually realistic representation of the apparatus. (b) Biprism experiment from the Optics Laboratory (OpticsLab), presented in a more schematic style.}
        \end{figure}

        \autoref{fig:lab_interface} presents two example interfaces from the openphys.in platform, showcasing the range of design styles used across different experiments. Some interfaces, such as the Young's Modulus simulation, closely mimic the appearance of real laboratory apparatus, while others, like the Biprism simulation, adopt a more schematic and abstract layout. 
        
        The user interface is designed to be intuitive and visually representative of real laboratory apparatus. Interactive controls such as sliders, dials, and input fields allow for precise adjustment of experimental parameters (as seen in \autoref{fig:lab_interface}). Each experiment is accompanied by a theoretical overview and step-by-step procedural guidance to reinforce conceptual learning. Students can record their experimental data directly from the platform like real lab experiments, with no built-in calculation or automated data analysis tools provided. This approach encourages students to engage actively with the data, performing calculations and analysis independently as they would in a physical laboratory setting.
        
        The platform is usable from desktops, laptops and tablets; and incorporates accessibility features such as keyboard navigation and clear visual cues. Its lightweight architecture ensures high availability and ease of maintenance, supporting broad adoption across educational contexts.

        \subsection{Modularity and Extensibility}

        The modular structure of \texttt{openphys.in} allows for straightforward expansion and adaptation. New experiments or features can be added with minimal changes to the existing codebase, supporting ongoing curriculum development and customization for different institutional needs. The platform's open, standards-based architecture facilitates interoperability and future integration with learning management systems or other educational tools \citep{semiopenplatform,gigaspacesopenplatform}.
        
        Each simulation is encapsulated within its own folder, containing all necessary files and resources. This directory-based modularization ensures clean separation between experiments. For example, the Young's Modulus simulation resides in a folder named \texttt{young/} and other experiments follow the same convention.
        
        A typical experiment module consists of:
        \begin{itemize}
            \item \texttt{main.html} - the primary file defining the user interface layout and embedding the majority of scripts and styling.
            \item \texttt{stl.css} - an additional external stylesheet used for specific layout or appearance refinements.
            \item \texttt{rsrc/} - a resource folder that contains experiment-specific graphics and interface images.
            \item Optional JavaScript file - some experiments include an additional \texttt{.js} file for modular scripting or handling more complex calculations.
        \end{itemize}
        
        Notably, most styling (CSS) and scripting (JavaScript) are embedded directly within \texttt{main.html}, including page loader animations, responsive layout behavior and input controls. This design minimizes dependencies and simplifies static deployment, which is particularly advantageous for platforms hosted on services like GitHub\textsuperscript{\textregistered{}} Pages.
        
        This structure allows rapid duplication and customization. An instructor can replicate an entire experiment directory, modify a few parameters and deploy a new version with minimal technical effort. The approach also facilitates collaboration and version control, enabling institutions to adapt and extend the platform according to their specific curricular requirements.
        
\begin{figure}[htbp]
            \centering
            \includegraphics[width=0.4\linewidth]{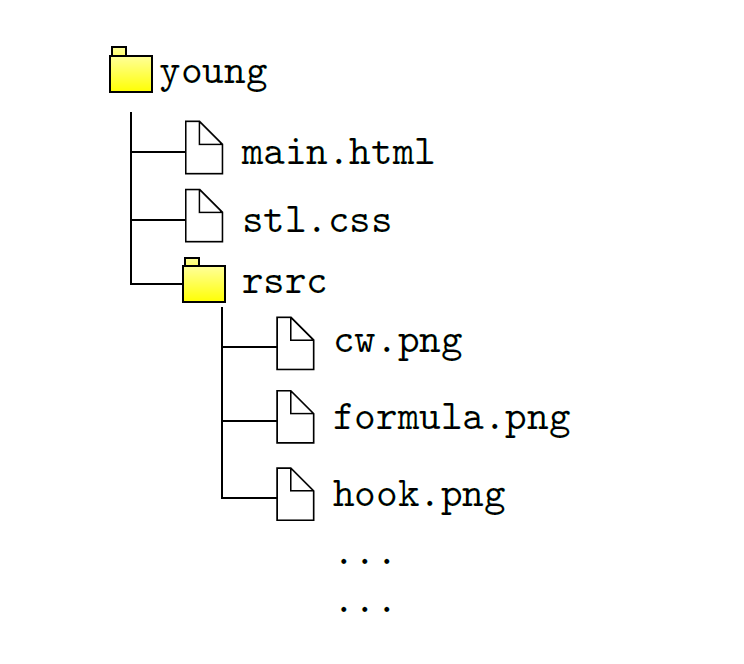}
            \caption{\label{fig:exp-structure} Example directory structure of an individual experiment on \texttt{openphys.in}. Shown here is the folder layout for the Young's Modulus simulation, which includes an HTML file for the interface, a CSS file for styling and a resource folder (\texttt{rsrc/}) containing graphical elements. Similar modular organization is followed for all other experiments on the platform.}
        \end{figure}

        Figure~\ref{fig:exp-structure} illustrates the folder hierarchy of a typical experiment module on the platform. The example shown corresponds to the Young's Modulus simulation, which follows a modular and self-contained structure. Similar layouts are adopted across all other experiments, enabling consistent development, easy duplication, and straightforward customization within the platform.
        
        \subsection{Deployment and Implementation}

        To ensure broad accessibility and ease of use, the \texttt{openphys.in} platform was deployed using GitHub\textsuperscript{\textregistered{}} Pages, a widely used open-source static website hosting service \citep{githubpages}. This approach leverages the platform's lightweight, client-side architecture—built entirely with HTML, CSS, and JavaScript—allowing the site to be hosted without the need for any backend server or runtime environment \citep{everhourgithub,kennyyipgithub}. The platform files are publicly available in a GitHub\textsuperscript{\textregistered{}} repository, supporting transparency and potential community contributions.
        
        A custom domain, \texttt{openphys.in}, was registered and configured to point to the GitHub\textsuperscript{\textregistered{}} Pages deployment, providing a professional and memorable access point for users. While GitHub\textsuperscript{\textregistered{}} Pages hosting is free, the domain registration incurs a nominal cost.
        
        The platform was actively used at St. Xavier's College (Autonomous), Kolkata during the COVID-19 pandemic, when access to physical laboratories was restricted. This real-world deployment enabled uninterrupted laboratory instruction and provided students with hands-on experience in a virtual setting, validating the platform's practical utility and scalability in an academic environment.

    \section{\label{sec:res_method} Research Methodology}
        
        To evaluate the effectiveness of the \texttt{openphys.in} online laboratory platform, we adopted a mixed-methods research design that combined quantitative and qualitative approaches. This methodology enabled us to assess both measurable learning outcomes and the nuanced experiences of students using the platform during remote learning at St. Xavier's College, Kolkata.
        
        \subsection{Study Context and Participants}
        
        The study was conducted with undergraduate physics students at St. Xavier's College, Kolkata, who utilized the platform as part of their laboratory coursework during the COVID-19 pandemic. Participation was voluntary, and students engaged with the virtual experiments prior to or alongside their physical laboratory sessions.
        
        \subsection{Student Feedback Instrument}
        
        A structured feedback form was developed to capture students' perceptions of the platform's usability, realism, and educational value. The questionnaire consisted of both closed and open-ended items, including:
        
        \begin{itemize}
            \item \textbf{Ease of Navigation and Use:} Rated on a 5-point Likert scale.
            \item \textbf{Perceived Realism:} Students rated how realistic the simulations felt compared to real experiments (1 = Not at all realistic, 5 = Very realistic).
            \item \textbf{Helpfulness for Learning:} Students rated how much the simulations helped them understand experimental concepts before performing them in a physical lab.
            \item \textbf{Confidence Building:} Students rated their confidence in performing physical experiments after using the simulations.
            \item \textbf{Efficiency in Physical Labs:} Students indicated if the simulations helped them set up and execute physical experiments more efficiently.
            \item \textbf{Comparative Benefit:} Students assessed whether having access to an online pre-lab simulation was beneficial compared to only physical lab learning.
            \item \textbf{Key Features:} Students selected the most useful features (e.g., realistic interface, interactive controls, data collection tools, self-paced learning) and could specify others.
            \item \textbf{Challenges Faced:} Open-ended responses on difficulties encountered.
            \item \textbf{Comparative Understanding:} Open-ended responses on how the simulation improved or did not improve their understanding compared to the physical lab.
            \item \textbf{Overall Recommendation:} Students indicated whether they would recommend the platform to others (Yes/No/Maybe).
        \end{itemize}
        
        \subsection{Performance Assessment}
        
        To quantitatively measure learning outcomes, students completed assessments before (pre-test) and after (post-test) using the platform. The pre-test evaluated their baseline understanding of experimental concepts, while the post-test assessed improvements following engagement with the virtual labs. The difference in scores was analyzed to determine the platform's impact on conceptual understanding and experimental readiness.
        
        \subsection{Qualitative Interviews}
        
        To gain deeper insights, structured interviews were conducted with a subset of student participants. These interviews explored the usability of the platform, challenges faced, and suggestions for improvement. Faculty members were also interviewed to gather feedback on the pedagogical effectiveness and integration of the platform within the curriculum.
        
        \subsection{Data Analysis}
        
        Quantitative data from Likert-scale survey items and pre/post-tests were analyzed using descriptive statistics, with results reported as percentages to facilitate clear interpretation and comparison~\citep{AERA2006}. Qualitative data from open-ended survey responses and interviews were thematically analyzed to identify recurring themes, challenges, and recommendations for future development.
        
        This comprehensive methodology provided a robust evaluation of \texttt{openphys.in}, capturing both objective learning outcomes and subjective user experiences in an authentic educational context.

    \section{\label{sec:result_discus} Results and Discussion}
    
        Analysis of the collected survey and interview data provided a comprehensive view of students' experiences with the online simulation platform. \autoref{fig:surveybar} presents a summary of the survey responses across key aspects, including perceived benefit, efficiency, confidence, understanding, realism, and ease of use. The following sections summarize the quantitative outcomes for each survey dimension and highlight recurring themes from the qualitative feedback, offering insights into both the strengths and areas for enhancement of the platform.
        \subsection{Ease of Use}
        
        \textbf{Survey Question:} How easy was it to navigate and use the online simulation platform? (1 = Very Difficult, 5 = Very Easy)
        
        \textbf{Results:} 95.16\% of students rated the platform as ``Easy'' or ``Very Easy'' (4 or 5).
        
        \textbf{Inference:} The vast majority of students found the platform highly user-friendly, indicating an intuitive and accessible interface.
        
        \subsection{Realism of Simulated Experiments}
        
        \textbf{Survey Question:} How realistic did the simulated experiments feel compared to real experiments? (1 = Not at all Realistic, 5 = Very Realistic)
        
        \textbf{Results:} 96.77\% of students rated the realism positively (4 or 5).
        
        \textbf{Inference:} Most students felt that the simulations closely resembled physical experiments. Some qualitative responses suggested further improvements in visual detail and interactivity.
        
        \subsection{Conceptual Understanding}
        
        \textbf{Survey Question:} To what extent did the simulations help you understand the experimental concepts before performing them in a physical lab? (1 = Not Helpful, 5 = Very Helpful)
        
        \textbf{Results:} 100.00\% of students found the simulations helpful (4 or 5), with 69.35\% selecting ``Very Helpful'' and 30.65\% selecting ``Helpful''.
        
        \textbf{Inference:} The simulations effectively prepared students for real experiments, with many noting that the pre-lab experience reduced confusion during hands-on work.
        
        \subsection{Confidence Boost}
        
        \textbf{Survey Question:} After using the online simulations, did you feel more confident performing the experiments in the physical lab? (1 = Less Confident, 5 = Very Confident)
        
        \textbf{Results:} 100.00\% of students reported increased confidence, with 64.52\% selecting ``Very Confident'' and 35.48\% selecting ``Confident''.
        
        \textbf{Inference:} The simulations helped students feel more prepared and less anxious when handling real equipment.
        
        \subsection{Efficiency in Experiment Execution}
        
        \textbf{Survey Question:} Did the online simulation help you set up and execute the physical experiment more efficiently? (1 = Not Beneficial, 5 = Very Beneficial)
        
        \textbf{Results:} 100.00\% of students responded positively, with 54.84\% selecting ``Very Beneficial'' and 45.16\% selecting ``Beneficial''.
        
        \textbf{Inference:} The platform successfully optimized students' preparation time and efficiency in the lab. Some students suggested that step-by-step guidance could further enhance efficiency.
        
        \subsection{Perceived Benefit of Online Pre-Lab Simulations}
        
        \textbf{Survey Question:} Compared to learning only in a physical lab, do you think having access to an online pre-lab simulation is beneficial? (1 = Not Beneficial, 5 = Very Beneficial)
        
        \textbf{Results:} 100.00\% of students responded positively (4 or 5), with 91.94\% selecting ``Very Beneficial'' and 8.06\% selecting ``Beneficial''.
        
        \textbf{Inference:} The results strongly indicate that online pre-lab simulations provide significant advantages in preparing students for physical experiments.
        
        \subsection{Recommendation as a Pre-Lab Practice Tool}
        
        \textbf{Survey Question:} Would you recommend using the online simulations as a regular pre-lab practice tool? (Options: Yes / No / Maybe)
        
        \textbf{Results:} 100.00\% of students would recommend the platform as a regular pre-lab practice tool.
        
        \textbf{Inference:} All students supported integrating the platform as a standard pre-lab practice tool. Open-ended responses suggested expanding the platform to more experiments.
        
        \subsection{Most Useful Features}

        The students selected the features they found most useful in the online simulation platform. The results reveal a clear hierarchy of functionalities:
        
        \begin{itemize}
            \item \textbf{Realistic experimental interface:} 98.39\%
            \item \textbf{Ability to adjust parameters interactively:} 96.77\%
            \item \textbf{Self-paced learning flexibility:} 95.16\%
            \item \textbf{Data collection and analysis tools:} 87.10\%
        \end{itemize}
        
        The data show that the \emph{realistic experimental interface} was the most widely appreciated feature, closely followed by \emph{interactive parameter adjustment} and \emph{self-paced learning flexibility}. The high selection rates for these features underscore the importance of an authentic, hands-on experience and the ability for students to control their own pace of exploration. While \emph{data collection and analysis tools} were also highly valued, their slightly lower selection rate suggests that, although important, they are seen as a complement to the core interactive and immersive elements of the platform.
        
        This pattern highlights that students prioritize features that closely mimic real laboratory conditions and empower them to engage actively and independently with the experimental content. These findings provide clear guidance for further development: continued investment in interface realism and interactivity is likely to yield the greatest educational impact.
        
        \subsection{Qualitative Feedback and Challenges}
        
        Open-ended responses revealed that most students encountered no major difficulties, describing the experience as smooth and effective. However, several recurring suggestions emerged:
        
        \begin{itemize}
            \item \textbf{Guided tutorial mode}: Requests for step-by-step instructions or a tutorial for beginners.
            \item \textbf{Enhanced data tools}: Desire for downloadable reports and built-in calculation or graphing features.
            \item \textbf{Improved visualization}: Suggestions for richer graphical representations and more immersive visual feedback.
            \item \textbf{Live feedback}: Requests for real-time textual cues or feedback during experiments.
            \item \textbf{More detailed explanations}: Calls for additional theoretical and procedural guidance within simulations.
        \end{itemize}
        
        A representative comment from the survey was:  
        \textit{``The GE Lab experiment simulation provides a simple and well-defined insight into performing an experiment and what exactly needs to be done. The only challenge is not being able to perform in a physical environment, but it certainly provides a precursor to performing the experiments in a physical lab.''}
        
        \subsection{Distribution of Responses}
        
\begin{figure*}[ht]
            \centering
            \includegraphics[width=0.9\linewidth]{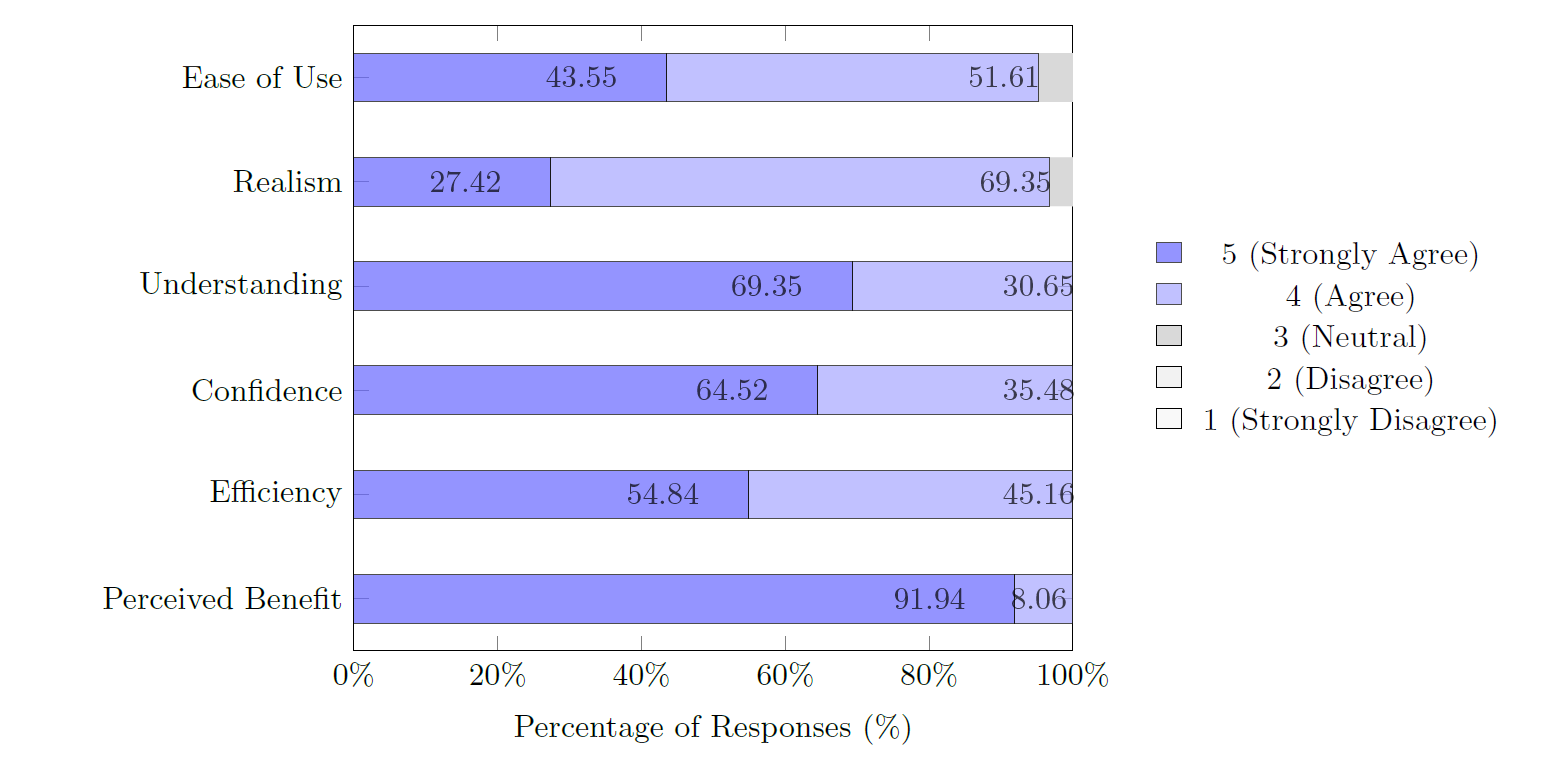}
            \caption{\label{fig:surveybar} Distribution of survey responses for key aspects of the online platform.}
        \end{figure*}

        Student feedback was categorized into six key aspects: perceived benefit, efficiency, confidence, understanding, realism, and ease of use. The majority of students responded positively across all categories, with most ratings falling within 4 (Agree) and 5 (Strongly Agree), indicating high satisfaction levels (see \autoref{fig:surveybar}).
        
        \begin{itemize}
            \item \textbf{Perceived Benefit}: 100.00\% of students responded positively (4 or 5), with 91.94\% selecting ``Very Beneficial'' (5) and 8.06\% selecting ``Beneficial'' (4).
            \item \textbf{Efficiency}: 100.00\% of students responded positively, with 54.84\% selecting option 5 and 45.16\% selecting option 4.
            \item \textbf{Confidence}: 100.00\% of students rated 4 or 5, indicating a boost in confidence.
            \item \textbf{Understanding}: 100.00\% of students found the simulations significantly helpful in understanding concepts, with 69.35\% selecting `very helpful' and 30.65\% selecting `helpful'.
            \item \textbf{Realism}: 96.77\% of students rated the realism positively (4 or 5).
            \item \textbf{Ease of Use}: 95.16\% of students rated it as ``Very Easy'' or ``Easy'' (4 or 5).
        \end{itemize}
        
        Overall, the findings confirm the platform's effectiveness in improving students' experimental skills, with minor improvements in realism and interactivity suggested for further development. 

    \section{\label{sec:comp_study} Comparative Analysis of Platform Strengths and Areas for Improvement}

        The comprehensive evaluation of student feedback reveals both clear strengths and actionable areas for improvement in the online simulation platform. This analysis is essential for guiding future enhancements and aligning the platform's development with best practices in simulation-based learning~\cite{frontiers2024, pmc2024, neovation2023}.
        
        \subsection{Platform Strengths}
        
        \textbf{1. High Perceived Benefit and Recommendation Rates:}  
        An overwhelming majority of students (91.94\% rating ``Strongly Agree'') believe that the platform is a valuable supplement to traditional laboratory learning. All respondents indicated they would recommend the simulations as a regular pre-lab practice tool, underscoring broad acceptance and perceived value.
        
        \textbf{2. Enhanced Conceptual Understanding and Confidence:}  
        The platform was particularly effective in improving students' conceptual understanding (69.35\% ``Strongly Agree'', 30.65\% ``Agree'') and boosting their confidence before performing physical experiments (64.52\% ``Strongly Agree'', 35.48\% ``Agree''). These outcomes align with the literature, which highlights simulation-based learning as a means to bridge theory and practice and foster deeper engagement~\cite{frontiers2024,sage2020}.
        
        \textbf{3. Efficiency and Usability:}  
        A combined 100\% of students rated the platform as efficient in helping them set up and execute physical experiments (54.84\% ``Strongly Agree'', 45.16\% ``Agree''). Ease of use was also highly rated, with 95.16\% selecting ``Very Easy'' or ``Easy.'' These results suggest that the platform's design supports both effective learning and user-friendly navigation.
        
        \textbf{4. Most Valued Features:}  
        Students identified the realistic experimental interface (98.39\%), interactive parameter adjustment (96.77\%), and self-paced learning flexibility (95.16\%) as the most useful features. These preferences reinforce the importance of authentic, interactive, and flexible learning environments in simulation-based education.
        
        \subsection{Areas for Improvement}
        
        \textbf{1. Realism of Simulations:}  
        While 96.77\% of students rated the simulations as ``realistic'' or ``highly realistic,'' only 27.42\% selected the highest rating. This indicates that, although the majority found the simulations sufficiently realistic, there is a notable opportunity to further enhance the fidelity, interactivity, and alignment with real-world laboratory conditions. Similar findings in recent studies recommend improvements in graphical fidelity, interface operability, and the inclusion of live operation videos to increase realism and engagement~\cite{pmc2024, neovation2023}.
        
        \textbf{2. User Interface and Accessibility:}  
        A small proportion of students (4.84\%) were neutral regarding ease of use, suggesting that onboarding, interface design, or accessibility could be further optimized. Recommendations from literature include refining toolbars, adding step-by-step instructions, and developing mobile-friendly versions to broaden accessibility and user satisfaction~\cite{pmc2024,neovation2023}.
        
        \textbf{3. Expanding Feature Set:}  
        Qualitative feedback and best practices highlight the potential value of additional features such as differentiated learning and assessment modes, smoother software operation, and enhanced data analysis tools~\cite{pmc2024, frontiers2024}. Incorporating these elements could further enrich the learning experience and better address diverse learner needs.
        
        \subsection{Opportunities for Future Development}
        
        In addition to addressing current areas for improvement, several opportunities have been identified to further enhance the platform's educational impact and accessibility:
        
        \begin{itemize}
            \item \textbf{Integrate live experiment videos:} Embedding high-quality video demonstrations of real laboratory procedures can bridge the gap between virtual and physical experiences, providing students with richer context and a clearer understanding of experimental nuances.
            \item \textbf{Develop mobile-friendly version:} Optimizing the platform for mobile devices will increase accessibility, allowing students to engage with simulations anytime and anywhere, and supporting a wider range of learning environments.
            \item \textbf{Include adaptive learning pathways:} Implementing adaptive features that tailor content and feedback to individual student progress can enhance engagement, address diverse learning needs, and promote mastery of laboratory skills.
        \end{itemize}
        
        Pursuing these opportunities will not only address current student feedback but also position the platform at the forefront of innovation in simulation-based science education. These enhancements align with modern trends in digital learning and are expected to further improve student outcomes and satisfaction.

    \section{\label{sec:conclu} Summary and Conclusions}
        
        This study provides a comprehensive evaluation of an online simulation platform designed to enhance laboratory education. Quantitative and qualitative feedback from students demonstrates that the platform is highly effective in supporting pre-lab preparation, with nearly all respondents reporting improved conceptual understanding, increased confidence, and greater efficiency in executing physical experiments.
        
        Key strengths identified include the platform's realistic experimental interface, interactive parameter adjustment, and flexibility for self-paced learning. The vast majority of students rated these features as highly useful, and all participants indicated they would recommend the simulations as a regular part of laboratory instruction. These findings align with current best practices in simulation-based education, which emphasize authenticity, interactivity, and accessibility as drivers of student engagement and learning outcomes~\cite{frontiers2024, pmc2024, neovation2023}.
        
        Areas for improvement were also identified, most notably in further enhancing the realism of simulations and refining the user interface to ensure universal ease of use. Opportunities for future development include integrating live experiment videos, developing a mobile-friendly version, and implementing adaptive learning pathways to address diverse learner needs.
        
        In conclusion, the online simulation platform has demonstrated substantial educational value and strong student acceptance. By building on its current strengths and addressing identified areas for enhancement, the platform can continue to advance laboratory education and serve as a model for effective digital learning tools in STEM disciplines \cite{pmcSTEM,sciencedirectSTEM}.
        
    \section*{Acknowledgements}

        The authors sincerely thank the faculty members of the Postgraduate and Research Department of Physics at St. Xavier's College (Autonomous), Kolkata for their invaluable guidance, encouragement and support throughout this project. Their expertise and constructive feedback during the laboratory sessions were instrumental in shaping the direction and quality of the research. The authors would like to extend special thanks to Dr. Tanaya Bhattacharyya for her insightful suggestions and encouragement, which played a significant role in refining the platform. The authors also acknowledge the students of the Department of Physics at St. Xavier's College (Autonomous), Kolkata for their enthusiastic participation and thoughtful feedback, which formed the foundation of the study's analysis and significantly contributed to its success.
	
    \bibliographystyle{apsrev4-2}
    \bibliography{references}
	
\end{document}